\documentstyle[sprocl]{article}

\def\Journal#1#2#3#4{{#1} {\bf #2}, #3 (#4)}
\def\NCA{\em Nuovo Cimento}

\def\NPB{{\em Nucl. Phys.} B}
\def\PLB{{\em Phys. Lett.}  B}

\def\PRD{{\em Phys. Rev.} D}
\def\ZPC{{\em Z. Phys.} C}

\def\ra{\rightarrow}
\def\ra\ppg{\pi^+\pi^-\gamma}

\def\vx{{\bf x}}
\def\bau{{\bf A}_1}
\def\bad{{\bf A}_2}

\def\FU{\eta_1}
\def\FD{\eta_2}

\def\be{\begin{equation}}
\def\ee{\end{equation}}
\def\bea{\begin{eqnarray}}
\def\eea{\end{eqnarray}}

\begin{document}

\title{SCHR\"ODINGER FUNCTIONAL AND QUANTIZATION OF
GAUGE THEORIES IN THE TEMPORAL GAUGE}

\author{G. C. ROSSI}

\address{Dipartimento di Fisica, Universit\`a di Roma ``Tor Vergata"\\
INFN - Sezione di Roma 2 \\
Via della Ricerca Scientifica - 00133 Rome - ITALY\\
E-mail: rossig@roma2.infn.it}

\maketitle
\abstracts{In the language of Feynman path integrals the quantization 
of gauge theories is most easily carried out with the help of the
Schr\"odinger Functional (SF). Within this formalism the essentially
unique gauge fixing condition is $A_{\circ} = 0$ (temporal gauge),
as any other rotationally invariant gauge choice can be shown to be 
functionally equivalent to the former. In the temporal gauge 
Gauss' law is automatically implemented as a constraint on the states.
States not annihilated by the Gauss operator describe the situation in
which external (infinitely heavy) colour sources interact with the gauge
field. The SF in the presence of an arbitrary distribution of external
colour sources can be expressed in an elegant and concise way.}

1. $-$ In the framework of the Feynman path integrals~\cite{FH} a key
role is played by the Schr\"odinger Functional (SF), $K(\FD,T_2;\FU,T_1)$
= $K(\FD,\FU;T_2-T_1)$, which represents the probability amplitude  of finding
the fundamental fields of the theory, $\{\eta({\bf x},t)\}$, in the
configuration $\eta_2 ({\bf x})$ at time  $T_2$, if they were in the
configuration  $\eta_1 ({\bf x})$ at time $T_1$. $K(\FD,\FU;T_2-T_1)$ is
the matrix element of the time  evolution operator between eigenstates of the
field operator, in the  Schr\"odinger representation~\cite{S}. In formulae
we would schematically write ($T=T_2-T_1$)  
\begin{eqnarray}
K(\FD,\FU;T) &=& \langle \FD |\mbox{e}^{-i{\cal{H}}T/\hbar} |\FU
\rangle\nonumber\\
\hat{\eta}({\bf x},t_{\circ}) | \eta_{I}\rangle &=& 
\eta_{I}({\bf x})|\eta_{I}\rangle \qquad I=1,2\label{eq:ONE}
\end{eqnarray} 
where the ``hat" on $\hat{\eta}({\bf x},t_{\circ})$ has been introduced to 
make clear in the last equation the distinction between field operators and 
their corresponding eigenvalues, $\eta_{I}({\bf x})$. $t_{\circ}$ is a fixed,  
arbitrary time, at which the Schr\"odinger representation is defined.
The knowledge of $K(\FD,\FU;T)$ is sufficient to compute any physical quantity,
as seen from the formal expansion (spectral representation) 
\begin{equation}
K(\FD,\FU;T) =
\langle\FD|\mbox{e}^{-i{\cal{H}}T/\hbar} |\FU\rangle
=\sum_{k}\mbox{e}^{-iE_{k}T/\hbar}\Psi_{k}(\FD) 
\Psi_{k}^{\star}(\FU)
\label{eq:SREP}
\end{equation}
where the state functionals
$\Psi_{k}(\eta)=\langle\eta|\Psi_{k}\rangle$  are eigenstates of the
Hamiltonian, ${\cal{H}}$, of the system, ${\cal{H}}|\Psi_{k}\rangle$ =
$E_{k}|\Psi_{k}\rangle$.

In this talk I wish to illustrate the quantization of a gauge theory in the
gauge $A_{\circ}=0$, using the notion of SF~\cite{RT1}. For lack of space
I will concentrate on the case of pure Yang-Mills (YM) theories and limit
the discussion to the topologically trivial sector of the theory. The 
procedure for including fermions is not entirely trivial and can be
found in~\cite{LMRY}. Topology is discussed in the third paper of
ref.~\cite{RT1}. For applications of the notion of SF to lattice QCD
see~\cite{L}.
\vskip .3cm 
2. $-$ The elegance of the principle of gauge invariance is
counterbalanced by the necessity of introducing the non-canonical variable
$A_{\circ}$. In computing physical quantities one is then confronted 
with the problem of eliminating it by the use of some gauge fixing condition. 
The simplest choice is to set $A_{\circ}=0$. For obvious reasons this gauge is
often referred to as the temporal gauge~\cite{W}.

Some nice features of the temporal gauge are the absence of ghosts, 
the fact that $A^{i}$ and $\Pi^{i}=\dot{A}^{i}$ ($i=1,2,3$) are canonically 
conjugate variables and the natural interpretation of the solutions of the
Euclidean field equations (instantons) as configurations interpolating between 
vacua (pure gauges) with winding number differing by one unit (tunneling).
In the $A_{\circ}=0$ gauge, however, Gauss' law is apparently lost. I will
show that Gauss' law is, in fact, naturally implemented as a constraint on the
states~\cite{RT1}. 

In the language of path integrals the SF of a YM theory,
$K_{_{YM}}(\bad,\bau;T)$, can be represented by the formal expression 
\begin{equation}
K_{_{YM}}(\bad,\bau;T) \sim 
\int^{\bad(\vx)}_{\bau(\vx)} \prod_{x}[d{\bf A}(x)
dA_{\circ}(x)]\, \exp(i\int_{T_1}^{T_2}\!{\cal{L}}_{_{YM}}[A_\mu]\,d{\bf x} dt)
\label{eq:KPI}
\end{equation}
where the functional integration is carried out over all field 
configurations such that ${\bf A}({\bf x},T_{1,2})$ = ${\bf A}_{1,2}({\bf x})$.
${\cal{L}}_{_{YM}}[A_\mu]=-\frac{1}{4}F^a_{\mu\nu}F^{a\mu\nu}$ is the YM
Lagrangian and $SU(N_c)$ is the gauge group. As $A_{\circ}$ is
the Lagrange multiplier enforcing Gauss' law, the integration over
$A_{\circ}$ in~(\ref{eq:KPI}) is extended to all times $T_1\le t \le T_2$
(including the boundaries).

Of course the integration in~(\ref{eq:KPI}) yields an infinite result as 
the integrand is invariant under local gauge transformations ($g_{\circ}$
is the YM coupling constant) 
\begin{equation}
\begin{array}{rcl}
&&A_{\mu}(x)\rightarrow A_{\mu}^{U_h}(x) = U_h^{\dagger}(x)A_{\mu}(x) U_h(x)
+\frac{i}{g_{\circ}} U_h^{\dagger}(x)\partial_{\mu} U_h(x)\\
&&U_h(x)=\exp[ig_{\circ}h(x)] \label{eq:AGAUGE}\\
&&A_{\mu}(x)-A_{\mu}^{U_h}(x)\simeq D_{\mu}(A)h(x)=\partial_\mu h(x)-i g_{\circ}[A_\mu(x),h(x)]
\end{array}
\end{equation}
To give a meaning to eq.~(\ref{eq:KPI}) we introduce in the path integral the identity
\begin{equation}
1=\Delta_{FP}[A_{\circ}]\int_{\cal{G}}\prod_{\vx}\prod_{t\in[T_1,T_2]}
{\cal{D}}h({\bf x},t) \delta(A_{\circ}^{U_h})
\label{eq:FP}
\end{equation}
where ${\cal{G}}$ is the group of topologically trivial gauge transformations 
and by $\prod {\cal{D}}h$ we mean the product of the Haar measures of  the gauge
group $SU(N_c)$ over all space-time points. $\Delta_{FP}[A_{\circ}]$ is the usual
Fadeev-Popov (FP) factor, which in this gauge only depends on $A_{\circ}$.
From eqs.~(\ref{eq:AGAUGE}) and~(\ref{eq:FP}) one obtains    
\begin{equation}
\begin{array}{rcl} 
&&K_{_{YM}}(\bad,\bau;T) \sim
\int \prod_{\vx}\prod_{t\in[T_1,T_2]} {\cal{D}}h({\bf x},t) \cdot
\label{eq:KPIFP1}\\&&
\cdot\int^{\bad(\vx)}_{\bau(\vx)} \prod_{x}[d{\bf A}(x)
dA_{\circ}(x)] \Delta_{FP}[A_{\circ}]\delta(A_{\circ}^{U_h})\, 
\exp(i\int_{T_1}^{T_2}{\cal{L}}_{_{YM}}[A_\mu]\,d{\bf x} dt) 
\nonumber 
\end{array}
\end{equation}
Putting $A_{\mu}^{U_h}=A_{\mu}^{'}$, the $A_{\circ}^{'}$ integration is
immediately carried out with the help of the $\delta$-function. Dropping the
field independent FP factor, one gets    
\begin{equation}
\begin{array}{rcl}
&&K_{_{YM}}(\bad,\bau;T) \sim 
\Big[\int \prod_{\vx}\prod_{t\in(T_1,T_2)} {\cal{D}}h({\bf x},t) 
\Big]\int_{{\cal{G}}_{\circ}} {\cal{D}}h_2({\bf x})\int_{{\cal{G}}_{\circ}} 
{\cal{D}}h_1({\bf x})\cdot \nonumber\\
&&\cdot\int^{\bad^{U_{h_2}(\vx)}}_{\bau^{U_{h_1}(\vx)}}
\prod_{x}d{\bf A}(x)\, \exp(i\int_{T_1}^{T_2}{\cal{L}}_{_{YM}}[{\bf A},
A_{\circ}=0]\,d{\bf x} dt) \nonumber
\end{array}
\end{equation}
where we have separated out the gauge integrations at the initial and
final time and introduced the definitions $U_h({\bf x},T_I)=U_{h_I}({\bf x})$, 
$I=1,2$. The $U_{h_{1,2}}$ integrations are extended over the group,
${\cal{G}}_{\circ}$, of the (topologically trivial) time independent gauge
transformations. As the integrand does not depend on the gauge transformations
at times $t\in(T_1,T_2)$, one can drop the infinite gauge volume,
${\cal{V}}=\int \prod_{\vx}\prod_{t\in(T_1,T_2)} {\cal{D}}h({\bf x},t)$, from
the previous equation and write 
\begin{equation}
\begin{array}{rlc} 
&&K_{_{YM}}(\bad,\bau;T) = 
\int_{{\cal G}_{\circ}} {\cal{D}}h_1({\bf x})\int_{{\cal G}_{\circ}} 
{\cal{D}}h_2({\bf x}) \,\widetilde{K}_{_{YM}}(\bad^{U_{h_2}},\bau^{U_{h_1}};T)
\label{eq:K} \\
&&\widetilde{K}_{_{YM}}(\bad,\bau;T)=\int^{\bad ({\bf x})}_{\bau ({\bf x})} 
\prod_{x} d{\bf A}(x)\,
\exp(i\int_{T_1}^{T_2}{\cal{L}}_{_{YM}}[{\bf A}, A_{\circ}=0]\,d{\bf x} dt)
\label{eq:KTILDE}
\end{array} 
\end{equation}

It should be immediately noted that, as a consequence of the invariance
$\widetilde{K}_{_{YM}}(\bad^V,\bau^V;T)$ =
$\widetilde{K}_{_{YM}}(\bad,\bau;T)$, $V\in{\cal{G}_{\circ}}$, the integrand
in the first of eqs.~(\ref{eq:K}) only depends on $U_{h_2}U^{\dagger}_{h_1}$.
Consequently one of the two gauge integrations must be dropped in order to
get a finite answer~\cite{RT2}, leading to the final formula  
\begin{equation}
K_{_{YM}}^{A_{\circ}}(\bad,\bau;T) =  \int_{{\cal G}_{\circ}} {\cal{D}}h({\bf x})
\,\widetilde{K}_{_{YM}}(\bad^{U_{h}},\bau;T)
\label{eq:KFIN} 
\end{equation}
From this equation one can construct a perfectly well defined perturbative
expansion with no singularity whatsoever in the gluon propagator~\cite{RT1} and
prove that no problems arise with the time exponentiation of the Wilson
loop~\cite{CCM}.

To understand the physical meaning of the
gauge ``projections" in the above formulae it is more convenient to refer our
considerations to  eqs.~(\ref{eq:K}). Inserting the spectral decomposition of 
$\widetilde{K}_{_{YM}}(\bad,\bau;T)$,
\begin{equation}
\widetilde{K}_{_{YM}}(\bad,\bau;T) =
\langle\bad|\mbox{e}^{-i{\cal{H}}_{YM} T/\hbar} |\bau\rangle=
\sum_{\gamma}\mbox{e}^{-iE_{\gamma}T/\hbar}\Phi_{\gamma}(\bad) 
\Phi_{\gamma}^{\star}(\bau)
\label{eq:KHAT}
\end{equation}
\begin{equation}
{\cal{H}}_{_{YM}} \Phi_{\gamma}({\bf A})=E_{\gamma}\Phi_{\gamma}({\bf A})
\qquad
{\cal{H}}_{_{YM}}=\frac{1}{2}\int d{\bf x}\Big[-\frac{\delta^2}{\delta
A^a_i\delta A^a_i}+ \frac{1}{2} F^a_{ij}F^a_{ij}\Big]
\label{eq:HC}
\end{equation}
in~(\ref{eq:K}), we get
\begin{equation}
\begin{array}{rcl}&&
K_{_{YM}}(\bad,\bau;T) =
\sum_{\gamma}\mbox{e}^{-iE_{\gamma}T/\hbar} \int_{G_{\circ}} {\cal{D}}h_2
\Phi_{\gamma}(\bad^{U_{h_2}})  \int_{{\cal{G}}_{\circ}} {\cal{D}}h_1
\Phi_{\gamma}^{\star}(\bau^{U_{h_1}})\equiv\nonumber\\ && \label{eq:KHAF}\\
&&\equiv\sum_{k}\mbox{e}^{-iE_{k}T/\hbar}
\Psi_{k}(\bad) \Psi_{k}^{\star}(\bau) \,\Longrightarrow\, \Psi_{k}({\bf
A})=\int_{{\cal{G}}_{\circ}} {\cal{D}}h \Phi_{k}({\bf A}^{U_{h}})\nonumber
\end{array} \nonumber
\end{equation}

We see from eqs.~(\ref{eq:KHAF}) that the states appearing in $K_{_{YM}}$ are
gauge ``projections" of the states in $\widetilde{K}_{_{YM}}$, hence they are
invariant under ${\cal{G}_{\circ}}$. In fact from known properties of the
Haar measure, one immediately proves $\Psi_{k}({\bf A}^V)$ =
$\Psi_{k}({\bf A})$.  This means that the functionals $\Psi_{k}$
in~(\ref{eq:KHAF}) are physical states in the sense that, besides being
eigenstates of ${\cal{H}}$, they also satisfy Gauss' law: 
\begin{equation} 
0=\frac{\delta\Psi_{k}({\bf A}^{U_{h}})} {\delta h^a ({\bf x})} |_{h=0} = 
\int d{\bf y}\frac{\delta\Psi_{k}({\bf A})}{\delta A^b_i({\bf y})} 
\frac{\delta (A^b_i({\bf y})^{U_{h}})}{\delta h^a({\bf x})}|_{h=0}=
D^{ab}_i({\bf A}) \frac{\delta\Psi_{k}({\bf A})}{\delta A^b_i({\bf x})} 
\label{eq:GAUSS}
\end{equation}
The chain of equalities in eq.~(\ref{eq:GAUSS}) shows that the operator
$D^{ab}_i({\bf A}) \delta / \delta A^b_i({\bf x})$ is (up to a factor)
the generator of time independent gauge transformations.

It is important to note at this point that the scalar product of states obeying
Gauss' law must be defined by dividing out the volume of the group
${\cal{G}}_{\circ}$ (as was done in going from eqs.~(\ref{eq:K})
to~(\ref{eq:KFIN}))~\cite{RT2}. A practical way of doing it is to resort to the 
FP trick and write 
\begin{equation} 
(\Psi_{1},\Psi_{2}) = \int  d{\bf A} \,
\Psi^{\star}_{1}({\bf A}) \Psi_{2}({\bf A}) \Delta^{\cal{F}}_{FP}[{\bf A}] 
\delta({\cal{F}}({\bf A}))
\label{eq:SCAL}  
\end{equation}
where ${\cal{F}}({\bf A})$ = 0 is any (rotationally invariant) spatial gauge
condition (e.g.~$\nabla {\bf A}$ = 0) and 
$\Delta^{\cal{F}}_{FP}[{\bf A}]$ is the corresponding FP factor.
\vskip .3cm
3. $-$ We now want to prove the equivalence of the temporal gauge with any
other (rotationally invariant) spatial gauge. To be precise we want to
show that the SF in the gauge ${\cal{F}}({\bf A})=0$, call
it $K^{\cal{F}}_{_{YM}}$, is equal to the SF, $K_{_{YM}}$, given by
eqs.~(\ref{eq:KHAT}), when the latter is restricted to boundary fields
satisfying the gauge conditions ${\cal{F}}({\bf A}_{1,2})$ = $0$. 

To be concrete let us consider the Coulomb case, in which  
${\cal{F}}({\bf A})$ = $\nabla {\bf A}$. As it will be clear,
however, our proof is valid for any (rotationally invariant) gauge fixing
condition. In the Coulomb gauge the SF, $K^{C}_{_{YM}}$, takes the form
\begin{equation}
\begin{array}{rcl}
K^{C}_{_{YM}}(\bad,\bau;T)&=& 
\int^{\bad}_{\bau} \prod_{x} dA_{\mu}(x) \Delta^{C}_{FP}[{\bf A}] \delta 
(\nabla {\bf A})\, \exp(i\int_{T_1}^{T_2}{\cal{L}}_{_{YM}}[A_\mu]\,d{\bf x} dt)\\
\\ 1&=&\Delta^{C}_{FP}[{\bf A}]\int_{\cal{G}}\prod_{\vx}\prod_{t\in(T_1,T_2)}
{\cal{D}}h({\bf x},t) \,\delta(\nabla {\bf A}^{U_h})
\label{eq:FPCOUL}
\end{array}
\end{equation}
where the second equation defines the FP factor, $\Delta^{C}_{FP}$. No
functional gauge integrations at $t=T_2$ and $t=T_1$ appear in
the above formulae as in the Coulomb gauge the boundary fields necessarily obey
the gauge conditions $\nabla {\bf A}_{1,2}=0$.

We now perform the change of variables ($A_{\circ}, {\bf A}$)
$\rightarrow$ ($h, {\bf A}^{\prime}$) defined by
\begin{equation}
\begin{array}{rcl}
&&{\bf A}^{\prime}(\vx,t)={\bf A}^{U_h}(\vx,t)= U^{\dagger}_h(\vx,t) {\bf A}(\vx,t)
U_h(\vx,t)+\frac{i}{g_{\circ}} U^{\dagger}_h(\vx,t) {\nabla }U_h(\vx,t)\\
&&A_{\circ}^{\prime}(\vx,t)\equiv U^{\dagger}_h(\vx,t) A_{\circ}(\vx,t) 
U_h(\vx,t)+\frac{i}{g_{\circ}} U^{\dagger}_h(\vx,t) \partial_{\circ} U_h(\vx,t) =0 
\end{array}
\label{eq:HCOV}
\end{equation}
for all $t\in[T_1,T_2]$. Solving the second of eqs.~(\ref{eq:HCOV}) 
for $U_h$, one gets 
$U_h(\vx,t)$ = $T\exp (ig_{\circ}\int_{T_1}^t A_{\circ}(\vx,\tau)d\tau)
\,U_h(\vx,T_1)$, from which one computes the integration measure in the new
variables to be $\delta {\bf A}\delta A_{\circ}$ = $\delta {\bf
A}^{\prime}J(h)\delta h$ = $\delta {\bf A}^{\prime} {\cal{D}} h$.
The appearance of the Haar measure in the last equality follows 
upon observing that
\begin{equation}
J(h)=det\frac{\delta A^a_{\circ}}{\delta
h^b}=det\frac{\mbox{e}^{i\chi}-1}{i\chi}, \quad \chi_{ab}=if^{abc}h^c
\label{eq:JH}
\end{equation}
Since from eqs.~(\ref{eq:HCOV}) one gets ${\cal{L}}_{_{YM}}[A_\mu]$ =
${\cal{L}}_{_{YM}}[A_\mu^{U_{h}}]$ = 
${\cal{L}}_{_{YM}}[{\bf A}^{\prime},A^{\prime}_{\circ}=0]$,
$K^{C}_{_{YM}}$ (see eqs.~(\ref{eq:FPCOUL})) can be written in the form  
\begin{equation}
\begin{array}{rcl} 
&&K^{C}_{_{YM}}(\bad,\bau;T) = 
\int \prod_{\vx}\prod_{t\in[T_1,T_2]} {\cal{D}}h({\bf x},t)
\int^{{\bad}^{U_{h}(\vx,T_2)}}_{{\bau}^{U_{h}(\vx,T_1)}}\cdot \\
&&\\&&\cdot\prod_x d{\bf A}^{\prime}(x)
\Delta^{C}_{FP}[{\bf A}^{\prime}] \delta  (\nabla {\bf
A}^{{\prime}U^{\dagger}_h})\, \exp(i\int_{T_1}^{T_2}{\cal{L}}_{_{YM}}[{\bf
A}^{\prime},A^{\prime}_{\circ}=0]\,d{\bf x} dt)\nonumber
\end{array}
\label{eq:KPCOUL}
\end{equation}
Exploiting backwards the definition of $\Delta^{C}_{FP}$ given in
eq.~(\ref{eq:FPCOUL}), we can trivially perform the gauge integrations for all
$T_1<t<T_2$. In this way the whole Coulomb gauge fixing term gets replaced by
unity. Only the initial and final gauge integrations are left out: it is then
immediately seen that formula~(\ref{eq:KPCOUL}) coincides with the expression
given by eqs.~(\ref{eq:K}), as announced.
\vskip .3cm
4. $-$ External (infinitely heavy) colour sources can be elegantly
introduced in the present formalism by noticing that from the invariance
property of $\widetilde{K}_{_{YM}}$ 
\begin{equation}  
\sum_{\gamma}\mbox{e}^{-iE_{\gamma}T/\hbar}\Phi_{\gamma}(\bad^V) 
\Phi_{\gamma}^{\star}(\bau^V) =
\sum_{\gamma}\mbox{e}^{-iE_{\gamma}T/\hbar}\Phi_{\gamma}(\bad) 
\Phi_{\gamma}^{\star}(\bau),\quad V\in{\cal{G}_{\circ}} \label{eq:KTINV}
\end{equation} 
it follows that either {\bf i)} $\Phi_{\gamma}$ is not degenerate and consequently 
$\Phi_{\gamma}({\bf A}^V)$ = $\Phi_{\gamma}({\bf A})$, implying that
$\Phi_{\gamma}$ is actually a physical state annihilated by Gauss' law, or
{\bf ii)} there is some degeneracy and we have ($\gamma = n,s$) 
$\Phi_{n,s}({\bf A}^V)$ = $R_{s}^{s^{\prime}}(V) \Phi_{n,s^{\prime}}({\bf A})$,
where $R(V)$ is some unitary representation of the group ${\cal{G}}_{\circ}$.

It has been proved in the first paper of ref.~\cite{RT1} that all unitary
finite dimensional representations of ${\cal{G}}_{\circ}$ are of the form
$R(V_w)$ = $\otimes^p_{j=1}\exp[ig_{\circ} w^a({\bf y}_j) M^a_{R_j}]$,
where the $M^a_{R_j}$'s are the generators of $SU(N_c)$ in the
representation $R_j$. 

To understand the physical meaning of the situation described in
{\bf ii)} let us consider the simple case in which $R(V_w) =
\exp[ig_{\circ} w^a({\bf y}) M^a_{R}]$. Then a state transforming according to
the representation $R$, $\Phi_{n,s}^{R}$, will obey the equation 
\begin{equation}
D^{ab}_i({\bf A}) \frac{\delta \Phi^{R}_{n,s}({\bf A})}{\delta A^b_i({\bf x})}
=-ig_{\circ}[M^a_{R}]_{s}^{s^{\prime}}\Phi^{R}_{n,s^{\prime}}({\bf A})
\delta(\vx-{\bf y}) 
\label{eq:PHITR}
\end{equation} 
Eq.~(\ref{eq:PHITR}) is telling us that $\Phi^{R}_{n,s}$ is not annihilated by
the Gauss operator, but describes the dynamics of a point-like external 
(infinitely heavy) colour source, belonging to the representation $R$ and
located in space at the point ${\bf y}$, in interaction with the gauge
field .

From the point of view of this analysis the formulae given in eqs.~(\ref{eq:K})
should be interpreted as yielding the projection of $\widetilde{K}_{_{YM}}$ on
the trivial representation of ${\cal{G}}_{\circ}$. This simple observation gives
us a way to generalize eqs.~(\ref{eq:K}) to encompass the construction of the
SF in the presence of arbitrary external colour sources. To this end it is enough to 
project $\widetilde{K}_{_{YM}}$ on the appropriate (unitary) representation of
${\cal{G}}_{\circ}$. To illustrate this procedure in a simple, but physically relevant, 
case we give the formula for the SF when the gauge field is coupled to a pair of
$q\, {\overline{q}}$-like sources. If ${\vx}_q$ and ${\vx}_{\overline{q}}$ are
the locations of the two sources in space, one finds
$$
K_{_{YM}}^{q\,\overline{q}}(\bad,s_2,r_2;\bau,s_1,r_1;T)=
\int_{{\cal{G}}_{\circ}}\!{\cal{D}}h
\widetilde{K}_{_{YM}}(\bad^{U_h},\bau;T)[U_h(\vx_q)]^{s_1}_{s_2}
[U_h^{\dagger}(\vx_{\overline{q}})]^{r_2}_{r_1}$$
where now the state functionals contributing to $K_{_{YM}}^{q\,\overline{q}}$ 
have two extra indices, $s,r=1,\ldots,N_c$, corresponding to the colour degrees
of freedom of the sources.


\section*{References}
\begin{thebibliography}{99}

\bibitem{FH} R.P.~Feynman and A.R.~Hibbs, {\it{Quantum Mechanics and Path 
Integrals}} (McGraw-Hill Book Co., New York, 1965).

\bibitem{S} K.~Symanzik,\Journal{\NPB}{190}{1}{1981}.

\bibitem{RT1} G.C.~Rossi and M.~Testa, \Journal{\NPB}{163}{109}{1980};
\Journal{\NPB}{176}{477}{1980}; \Journal{\NPB}{237}{442}{1984}.

\bibitem{LMRY} G.C.~Rossi and K.~Yoshida, 
\Journal{\NCA}{11}{101}{1989};
J.P~Leroy, J.~Micheli, G.C.~Rossi and K.~Yoshida, 
\Journal{\ZPC}{48}{653}{1990}.

\bibitem{L} M.~L\"uscher, Les Houches Lectures, 1997 (hep-lat/9802029) and references 
therein; S.~Sint \Journal{\NPB}{421}{135}{1994}.

\bibitem{W} H.~Weyl, {\it{The theory of groups and quantum mechanics}}
(Dover, 1950).

\bibitem{RT2}G.C.~Rossi and M.~Testa,, \Journal{\PRD}{29}{2997}{1984}.

\bibitem{CCM} S.~Caracciolo, G.~Curci and P.~Menotti, \Journal{\PLB}{113}{311}
{1982}.

\end{thebibliography}
\end{document}